\documentclass{article}
\usepackage{hiph-art}

\newcommand{\Pythia}{\textsc{Pythia}}
\newcommand{\Jetset}{\textsc{Jetset}}
\newcommand{\Fritiof}{\textsc{Fritiof}}

\def\PLB{{\em Phys. Lett.} {\bf B}}
\def\PRC{{\em Phys. Rev.} {\bf C}}
\def\NPA{{\em Nucl. Phys.} {\bf A}}

\usepackage{graphicx}
\volnumber{19} \issuenumber{1} \edyear{2004}                             
\frompage{000} \topage{000}                                              
\recrevdate{7 February 2005}                                              
\def\rightmark{Hadron attenuation at HERMES and JLab}
\def\leftmark{T.~Falter, {\it et al.}}
 
\title{Hadron attenuation at HERMES and Jefferson Lab} 
\authors{ 
{T.~Falter, W.~Cassing, K.~Gallmeister and U.~Mosel$^1$ %
\index{Falter, T.} 
\index{Cassing, W.} 
\index{Gallmeister, K.}
\index{Mosel, U.}
}\\[2.812mm]
{\normalsize
\hspace*{-8pt}$^1$ Institut f\"ur Theoretische Physik,\\ 
Universit\"at Giessen, Germany\\[0.2ex] 
}}
 
\abstract{We investigate the attenuation of hadrons in deep inelastic
lepton-nucleus scattering in the kinematical regime of the HERMES and
Jefferson Lab experiments. The calculation is carried out in the
framework of a BUU transport model. Our results indicate a strong influence
of (pre)hadronic final state interactions on the observed multiplicity
ratios.}

 
\makeindex
\begin{document}
 
\maketitle

\section{Introduction}
\label{sec:intro}
In deep inelastic scattering experiments the reaction products 
hadronize long before they reach the detector. Thus by using elementary 
nucleon targets one cannot obtain information on the space-time
picture of hadronization. A simple estimate of 
the hadron formation proper time via the hadronic radius $r_h$ yields hadron 
formation lengths of the order $\gamma\cdot r_h$ in the laboratory frame. At high 
energies the Lorentz factor $\gamma$ leads to formation lengths
that may easily exceed typical nuclear dimensions. By using nuclear targets one,
therefore, has the unique possibility to investigate the final-state interactions 
(FSI) of the prehadronic system and to study the dynamics of the hadronization process.

In the recent past the HERMES collaboration has carried out an extensive study
of hadron production in deep inelastic lepton-nucleus scattering using 27.6 GeV 
and 12 GeV positron beams \cite{HERMESDIS,HERMESDIS_new}. The observed attenuation of hadrons 
--compared to a deuterium target-- has basically led to two different interpretations: The authors
of Refs.~\cite{ArleoWang} assume that hadronization occurs far outside the nucleus and that
the attenuation is caused by a partonic energy loss prior to hadronization.
On the other hand color neutral prehadrons might
form rather early after the initial deep inelastic scattering event and undergo
FSI on their way out of the nucleus \cite{Falteralt,Falterneu,Kopeliovich,Accardi}. The latter 
effect should be even more pronounced in the kinematical regime of the ongoing
Jefferson Lab experiment which uses a lower energy (5.4 GeV) electron beam \cite{Brooks}.

In Ref.~\cite{Falterneu} we have given a thorough theoretical investigation
of hadron attenuation in lepton-nucleus scattering at HERMES and EMC
energies in the framework of a semi-classical Boltzmann-Uehling-Uhlenbeck (BUU) transport model. 
In this work we apply our model (with the same parameter set) to lower energies,
i.e.~HERMES at 12 GeV beam energy and the kinematical regime of the Jefferson Lab experiment.
 
\section{Model}
\label{sec:model}  
In Refs.~\cite{Falteralt,Falterneu,EffeFalter} we have developed a method to combine a quantum
mechanical coherent treatment of nuclear shadowing --as observed in high energy
photonuclear reactions-- with an incoherent coupled-channel description of the 
(pre)hadronic FSI using a BUU transport model. This is achieved by splitting the 
electron-nucleus reaction into two parts. In the first step the high-energy virtual photon
interacts with a bound nucleon inside the nucleus and produces a final 
state which is determined by using the Lund Monte Carlo generators 
\Pythia{} and \Fritiof.
In addition we account for nuclear effects such as binding energies, 
Fermi motion, Pauli-blocking of final state nucleons and nuclear shadowing. 
In the second step the final state is propagated through the nucleus within 
the coupled-channel transport model.

\begin{figure}[tb]
\begin{center}
\includegraphics[scale=1.]{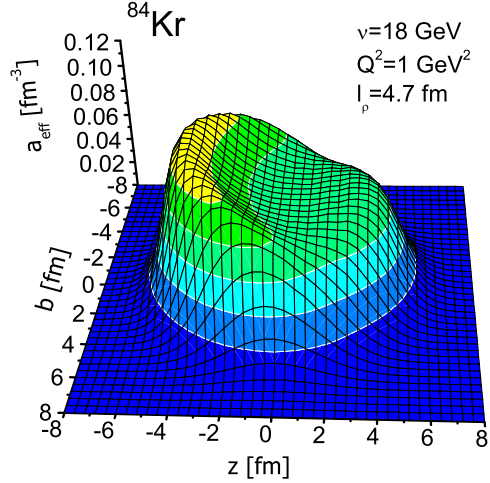}
\end{center}
\vspace{-1.5cm}
\caption[]{Profile function for shadowing. Shown is the probability
distribution for the interaction of an incoming photon (from left)
with given virtuality $Q^2$ and energy $\nu$ with nucleons in a $^{84}$Kr
nucleus.}
\label{fig:profile}
\end{figure}

The strength of the shadowing effect strongly depends on the coherence
lengths of the photon's hadronic fluctuations. The coherence length
can be understood as the distance that the virtual photon travels as
a hadronic fluctuation which has the quantum numbers of the photon, e.g.~a
vector meson. If the coherence length exceeds the mean-free path of the
hadronic fluctuation inside the nucleus, the photon-nucleus interaction
will get shadowed just like an ordinary hadron-induced nuclear reaction. 
In Fig.~\ref{fig:profile} we show the probability distribution for nucleons
inside a $^{84}$Kr nucleus to participate in the primary photon-nucleon 
interaction. The kinematics of the virtual photon with momentum along the
$z$-axis corresponds to that of a typical HERMES event. For such a photon the 
coherence length of the $\rho^0$-meson fluctuation is of the order of the 
nuclear radius and the nucleons on the front side of the $^{84}$Kr nucleus 
shadow the downstream nucleons.

The photon-nucleon interaction leads to the excitation of one or more
strings which fragment into color-neutral prehadrons due to the
creation of quark-antiquark pairs from the vacuum. As discussed in 
Ref.~\cite{Kopeliovich} the
production time of these prehadrons is very short. For simplicity
we set the production time to zero in our numerical realization.
These prehadrons are then propagated using our coupled-channel
transport theory.

After a formation time, which we assume to be a constant $\tau_f$
in the restframe of the hadron, the hadronic wave function has
built up and the reaction products propagate and interact like usual hadrons. The
prehadronic cross sections $\sigma^*$ during the formation time
are determined by a simple constituent quark model
\begin{eqnarray}
\label{eq:prehadrons} \sigma^*_\mathrm{prebaryon}&=
&\frac{n_\mathrm{org}}{3}\sigma_\mathrm{baryon} , \nonumber\\
    \sigma^*_\mathrm{premeson}&=&\frac{n_\mathrm{org}}{2}\sigma_\mathrm{meson}
    ,
\end{eqnarray}
where $n_\mathrm{org}$ denotes the number of (anti-)quarks in the
prehadron stemming from the beam or target. As a consequence the
prehadrons that solely contain (anti-)quarks produced from the
vacuum in the string fragmentation do not interact during
$\tau_f$. Using this recipe the total effective cross section of
the final state rises like in the approach of Ref.~\cite{Ciofi}
each time when a new hadron has formed.

The FSI of the reaction products are described within
our BUU transport model. The latter is based on a set of generalized 
transport equations for each particle species $i$,
\begin{equation}
\label{eq:BUU}
    \left(\frac{\partial}{\partial t}+\vec \nabla_{\vec p} H\vec \nabla_{\vec r}
    -\vec \nabla_{\vec r} H\vec \nabla_{\vec p}\right) F_i(\vec r,\vec p,\mu;t)=
    I_\mathrm{coll}(\{F_j\})
\end{equation}
where $H$ is a relativistic Hamilton function which contains a mean-field potential
in case of baryons. The transport equations (\ref{eq:BUU}) describe the time-evolution of the
spectral phase-space densities $F_i$ that are coupled
via the collision integral $I_\mathrm{coll}$. The latter
accounts for changes in the spectral phase-space density due to particle
creation and annihilation in binary collisions. This means that the 
final hadron in an electron-nucleus reaction does not necessarily need to be produced
in the primary virtual photon-nucleon interaction but can be created
later on via side-feeding in the FSI. We note that this (probabilistic) coupled-channel
treatment of the FSI goes far beyond the usual single-channel Glauber
approach.
\begin{figure}[tb]
\begin{center}
    \includegraphics[scale=1.3]{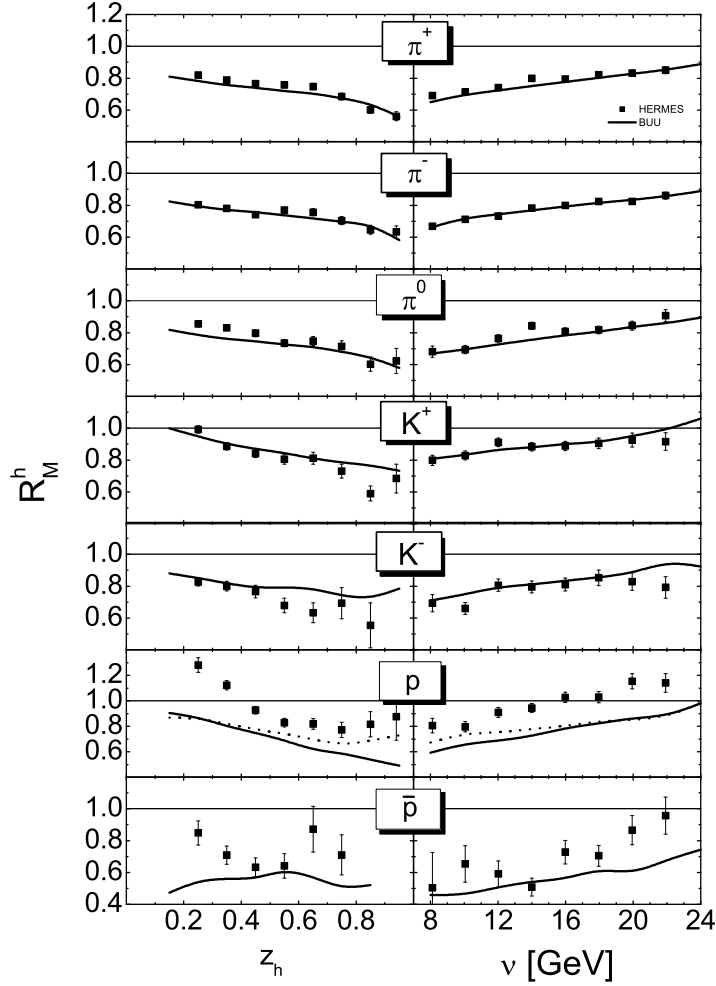}
\end{center}
  \caption[]{Multiplicity ratios of $\pi^{\pm,0}$, $K^\pm$, $p$ 
  and $\bar{p}$ for a $^{84}$Kr nucleus (when using a 27.6 GeV
  positron beam at HERMES) as a function 
  of the hadron energy fraction $z_h=E_h/\nu$ and the photon energy 
  $\nu$. The solid line represents the result of a simulation, 
  where we use the constituent quark concept (\ref{eq:prehadrons}) 
  for the prehadronic cross sections and a formation time 
  $\tau_f=0.5$ fm/$c$. The dotted line in the proton spectrum 
  indicates the result of a simulation where all $\gamma^*N$ 
  events are created by PYTHIA. The data are taken from 
  Ref.~\cite{HERMESDIS_new}.}
  \label{fig:Krid}
\end{figure}
  
\section{Results}\label{sec:results}
Due to our coupled channel-treatment of the FSI the (pre)hadrons 
might not only be absorbed in the
nuclear medium but can produce new particles in an inelastic interaction,
thereby shifting strength from the high to the low energy part of
the hadron spectrum. In addition our event-by-event simulation
allows us to account for all kinematic cuts and the
acceptance of the detector. In Ref.~\cite{Falterneu} we have
demonstrated that our calculations are in excellent agreement with
the experimental HERMES data \cite{HERMESDIS_new} taken on various
nuclear targets at a beam energy $E_\mathrm{beam}=27.6$ GeV if one
assumes a formation time $\tau_f=0.5$ fm/c for all hadron species.
As an example we show in Fig.~\ref{fig:Krid} the multiplicity ratios
\begin{equation}
\label{eq:multiplicity-ratio}
R_M^h(z_h,\nu)=\frac{\frac{N_h(z_h,\nu)}
{N_e(\nu)}\big|_A}{\frac{N_h(z_h,\nu)}{N_e(\nu)}\big|_D}
,
\end{equation}
for identified hadrons on a $^{84}$Kr target. 
Here $N_h$ is the yield of semi-inclusive hadrons in a given 
$(z_h,\nu)$-bin and $N_e$ the yield of inclusive 
deep inelastic scattering leptons in the same $\nu$-bin. 
For the deuterium target, i.e.~the nominator of 
Eq.~(\ref{eq:multiplicity-ratio}), we simply use the isospin 
averaged results of a proton and a neutron target. Thus in the 
case of deuterium we neglect the FSI of the produced hadrons 
and also the effect of shadowing and Fermi motion.

\begin{figure}[tb]
\begin{center}
\includegraphics[scale=1.15]{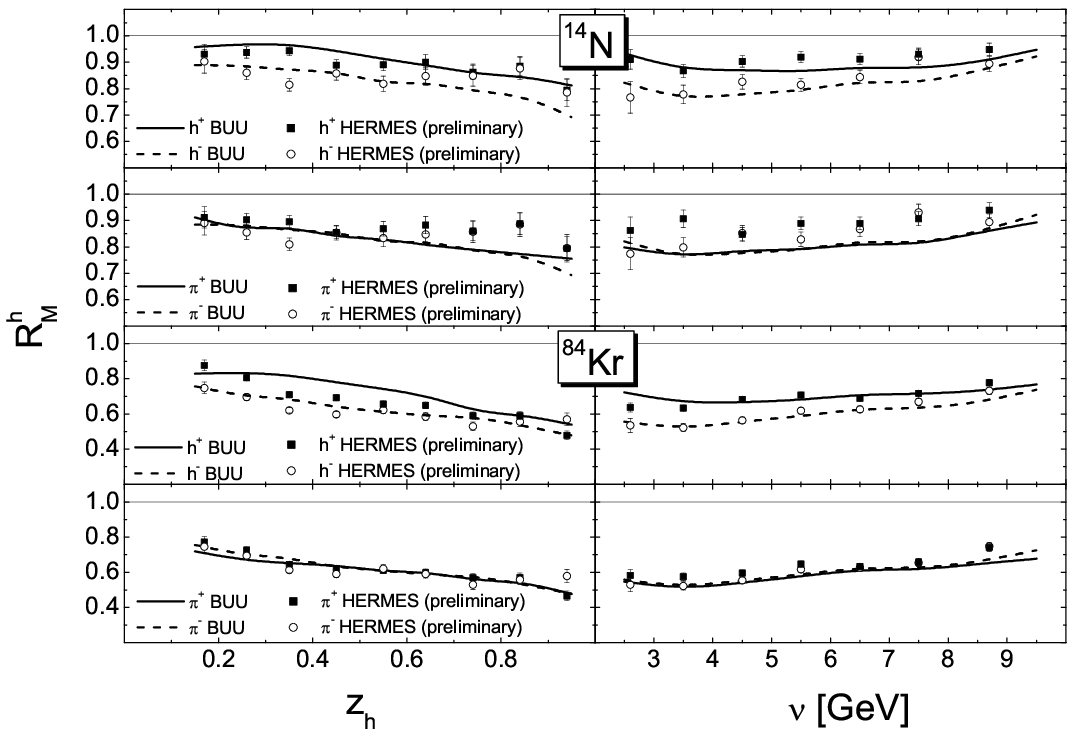}
\end{center}
\vspace{-0.5cm}
\caption[]{Calculated multiplicity ratio of
positively and negatively charged hadrons and pions for a $^{14}$N
and $^{84}$Kr target when using a 12 GeV positron beam at HERMES.
For the calculation we use the formation time $\tau_f=0.5$ fm/$c$
and the constituent-quark concept (\ref{eq:prehadrons}) for the
prehadronic cross sections. The data are taken from
Ref.~\cite{Nez04}.}
\label{fig:HERMES12}
\end{figure}

In Fig.~\ref{fig:HERMES12} we show that our approach is also
capable to describe the observed multiplicities of charged hadrons
in the HERMES experiment at $E_\mathrm{beam}=$12 GeV. This success
suggests that our model can also be applied for the electron beam
energies that will be used at Jefferson Lab.
\begin{figure}[htb]
\begin{center}
\includegraphics[scale=1.4]{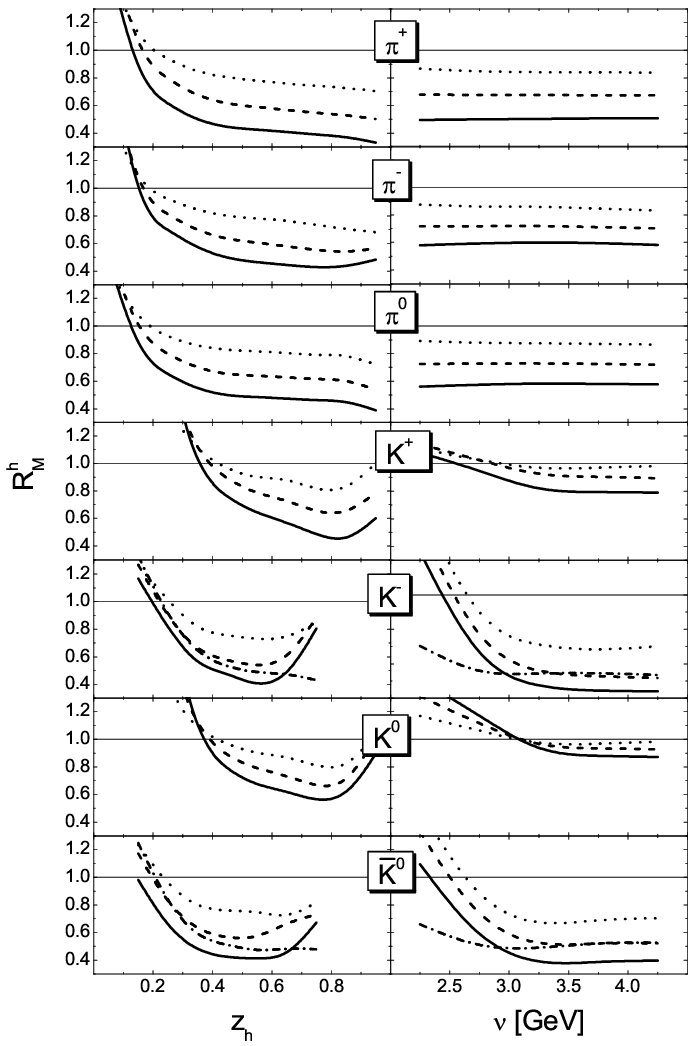}
\end{center}
\vspace{-0.5cm}
\caption[]{Calculated multiplicity ratio of identified $\pi^\pm$,
$\pi^0$, $K^\pm$, $K^0$ and $\bar{K}^0$ for $^{12}$C (dotted
lines), $^{56}$Fe (dashed lines) and $^{208}$Pb nuclei (solid
lines). The simulation has been done for a 5 GeV electron beam and
the CLAS detector. The dash-dotted line represents a calculation
for $^{56}$Fe without Fermi motion. In all calculations we use the
formation time $\tau_f=0.5$ fm/$c$ and the constituent-quark
concept (\ref{eq:prehadrons}) for the prehadronic cross sections.}
\label{fig:JLab}
\end{figure}

In Fig.~\ref{fig:JLab} we present our predictions for the
multiplicity ratios of identified hadrons at 5 GeV electron beam
energy. Besides the considerably lower beam energy used at
Jefferson Lab the major difference to the HERMES experiment is the
much larger geometrical acceptance of the CLAS detector. The
latter leads to an increased detection of low energy secondary
particles that are produced in the FSI and
that lead to a strong increase of the multiplicity ratio at low
fractional hadron energies $z_h=E_h/\nu$. In addition, the
relatively small average photon energy leads to a visible effect
of Fermi motion on the multiplicity ratio of more massive
particles. Since the virtual photon cannot produce antikaons
without an additional strange meson, e.g.~$\gamma^*N\to
K\bar{K}N$, the maximum fractional energy $z_h$ is limited to 0.9
for antikaons. Because of the energy distribution in the three
body final state and the finite virtuality of the photon -- set by
the kinematic cut $Q^2>1$ GeV$^2$ -- the maximum fractional energy
of antikaons is further reduced. As a result, the $z_h$ spectra
for $K^-$ and $\bar{K}^0$ in Fig.~\ref{fig:JLab} do not exceed
$z_h\approx0.8$. For the same reason the production of antikaons
with $z_h>0.2$ is reduced at the lower end of the photon spectrum.
The Fermi motion in the nucleus enhances the yield of antikaons in
these two extreme kinematic regions as can be seen by comparison
with the dash-dotted line in Fig.~\ref{fig:JLab} which represents
the result of a calculation for $^{54}$Fe where Fermi motion has
been neglected. Certainly, the kaons can be produced in a two-body
final state (e.g.~$\gamma^*N\to K\Lambda$), however, the
accompanying hyperon has a relatively large mass. Therefore,
similar effects, although less pronounced, show also up for the
kaons. Beside the effects of Fermi motion the multiplicity ratios
of kaons and antikaons show the same features as for higher
energies.

\section{Conclusions}\label{conclusions}
In this work we have presented a model that allows for a clean-cut
separation of the initial state interactions of a high energy
(virtual) photon --giving rise to nuclear shadowing-- and the nuclear 
FSI of the reaction products. This allows to apply
a coupled-channel transport code to perform realistic 
event-by-event simulations of high energy lepton-nucleus scattering
and to account for experimental cuts and detector efficiencies.

From our comparison with the experimental data we conclude that a large 
part of the observed hadron attenuation
at HERMES may be attributed to prehadronic FSI of the reaction products in the
nuclear environment. Furthermore, we also expect a strong effect of 
these prehadronic FSI at the considerably lower Jefferson Lab energies.

Our implemented space-time picture of hadronization is still simplistic
and needs improvements. In Ref.~\cite{neu} we have therefore
started to extract the four-dimensional prehadron production points
from the Lund fragmentation routine \Jetset{} in \Pythia. This information 
will be used in future transport simulations of deep inelastic 
lepton-nucleus scattering.

\section*{Acknowledgment}
This work has been supported by BMBF.

\vfill\eject
\end{document}